\documentclass[]{aastex631}

\usepackage{amsmath}
\usepackage{multirow}
\usepackage{ulem}

\begin{document}

\title{New Determination of the $^{14}$C(n, $\gamma$)$^{15}$C Reaction Rate and Its Astrophysical Implications}

\correspondingauthor{Yuchen Jiang}
\email{jiangyc18@tsinghua.org.cn}

\correspondingauthor{Jie Chen}
\email{chenjie@sustech.edu.cn}

\correspondingauthor{Toshitaka Kajino}
\email{kajino@buaa.edu.cn}

\correspondingauthor{Weiping Liu}
\email{liuwp@sustech.edu.cn}

\author[0000-0003-1032-8096]{Yuchen Jiang}
\affiliation{China Institute of Atomic Energy, Beijing, 102413, China}

\author[0009-0002-0124-9492]{Zhenyu He}
\affiliation{School of Physics, International Research Center for Big-Bang Cosmology and Element Genesis, Peng Huanwu Collaborative Center for Research and Education, Beihang University, Beijing, 100191, China}

\author[0000-0002-8965-1859]{Yudong Luo}
\affiliation{School of Physics, Peking University, and Kavli Institute for Astronomy and Astrophysics, Peking University, Beijing, 100871, China}

\author[0000-0003-3646-9356]{Wenyu Xin}
\affiliation{School of Physics and Astronomy, Beijing Normal University, Beijing, 100875, China}
\affiliation{Institute for Frontiers in Astronomy and Astrophysics,
Beijing Normal University, Beijing, 102206, China}

\author[0000-0003-4336-6726]{Jie Chen}
\affiliation{Department of Physics, Southern University of Science and Technology, Shenzhen, 518055, Guangdong, China}

\author[0000-0001-8333-0635]{Xinyue Li}
\affiliation{Institute of Modern Physics, Fudan University, Shanghai, 200433, China}

\author[0000-0003-4124-6034]{Yangping Shen}
\affiliation{China Institute of Atomic Energy, Beijing, 102413, China}

\author[0000-0002-4911-0847]{Bing Guo}
\affiliation{China Institute of Atomic Energy, Beijing, 102413, China}

\author[0000-0002-3856-1108]{Guo Li}
\affiliation{National Astronomical Observatories,Chinese Academy of Sciences, Beijing, 100020, China}

\author[0000-0002-2180-0758]{Danyang Pang}
\affiliation{School of Physics, Beihang University, Beijing 100191, People’s Republic of China}
\affiliation{Beijing Key Laboratory of Advanced Nuclear Materials and Physics, Beihang University, Beijing 100191, People’s Republic of China}

\author{Tianli Ma}
\affiliation{China Institute of Atomic Energy, Beijing, 102413, China}

\author[0000-0002-0857-5680]{Weike Nan}
\affiliation{Department of Physics, Southern University of Science and Technology, Shenzhen, 518055, Guangdong, China}

\author[0000-0003-1856-4173]{Toshitaka Kajino}
\affiliation{School of Physics, International Research Center for Big-Bang Cosmology and Element Genesis, Peng Huanwu Collaborative Center for Research and Education, Beihang University, Beijing, 100191, China}
\affiliation{National Astronomical Observatory of Japan, Mitaka, Tokyo, 181-8588, Japan}
\affiliation{Graduate School of Science, The University of Tokyo, Hongo, Tokyo, 113-033, Japan}

\author[0000-0003-3233-8260]{Weiping Liu}
\affiliation{Department of Physics, Southern University of Science and Technology, Shenzhen, 518055, Guangdong, China}
\affiliation{China Institute of Atomic Energy, Beijing, 102413, China}




\begin{abstract}
We present a novel experiment to investigate the spectroscopic factor of the $^{15}$C ground state for the first time using single-neutron $removal$ transfer reactions on $^{15}$C. Two consistent spectroscopic factors were derived from the (p, d) and (d, t) reactions, which were subsequently used to deduce the $^{14}$C(n, $\gamma$)$^{15}$C reaction cross section and the corresponding stellar reaction rate. A typical cross section of (3.89 $\pm$ 0.76) $\mu$b is determined at $E_\mathrm{_{c.m.}}$ = 23.3 keV. At the temperature range of 0.01-4 GK, our new reaction rate is 2.4-3.7 times higher than that of the first direct measurement and 20\%-25\% lower than that of the most recent direct measurement, respectively. Moreover, it is interesting that we can associate a long-standing nuclear structure issue, i.e., the so-called ``quenching'' effect, with this astrophysically relevant reaction. Finally, motivated by astrophysical interests of this reaction decades ago, implications of our new rate on several astrophysical problems are evaluated using state-of-the-art theoretical models. Our calculations demonstrate that the abundances of $^{14}$N and $^{15}$N can be enhanced in the inner regions of asymptotic giant branch (AGB) stars, though with minimal impact on the chemical compositions of the interstellar medium. In the inhomogeneous Big Bang nucleosynthesis, the updated reaction rate can lead to a $\sim 20\%$ variation in the final yields of $^{15}$N in neutron rich regions. For the $r$-process in the core-collapse supernovae, a slight difference of $\sim 0.2\%$ in the final abundances of heavy elements with $A > 90$ can be found by using our new rate.


\end{abstract}




\section{Introduction}~\label{sec:intro}

Radiative capture reactions are a prevalent type of reaction needed to model nucleosynthesis processes in nuclear astrophysics, where two nuclei collide and fuse into a heavier many-nucleon system with the emission of a photon. Among them, (n, $\gamma$) reactions are particularly distinctive due to their neutron-induced nature, where the reaction process is unaffected by the Coulomb barrier. It has been proved that such neutron capture reactions occur in the \textit{slow}~\citep{Lugaro_Annu.Rev.Nucl.Part.Sci._Process_2023, Kappeler_Rev.Mod.Phys._$s$_2011a} and \textit{rapid}~\citep{Cowan_Rev.Mod.Phys._Origin_2021} neutron-capture processes, called \textit{s}- and \textit{r}- processes, respectively, thus playing a key role in the synthesis of heavy elements and stellar evolution. However, direct measurement of such reactions within energy ranges of astrophysical interest remains challenging. Neutron capture reactions often involve short-lived isotopes, particularly those critical to the \textit{r}-process. It is extremely difficult or even unfeasible to make these unstable isotopes, as well as neutrons, into targets. Furthermore, neutron beam production and the corresponding measurement pose significant challenges compared to the measurement using charged particle beams from conventional accelerators, except for the extremely small cross sections of these reactions under astrophysical conditions. Although the development of facilities such as spallation neutron sources~\citep{Wei_Nucl.Instrum.MethodsPhys.Res.A_China_2009} and spallation-based neutron targets~\citep{Reifarth_Phys.Rev.Accel.Beams_Spallationbased_2017a} has recently introduced new possibilities, it remains a formidable task to achieve high-precision direct measurement of (n, $\gamma$) reaction cross sections.

To accurately characterize stellar evolution, reliable and well-constrained cross sections for these (n, $\gamma$) reactions are needed in theoretical nucleosynthesis calculations. Therefore, indirect techniques are vitally important to help us investigate these reactions, which are generally characterized by much larger cross sections. Several indirect techniques have been developed and applied to date, such as the transfer reaction method, charge symmetry method, surrogate reaction method, surrogate ratio method, Coulomb dissociation method and $\beta$-Oslo method (for some typical reviews in the past 5 years, see, e.g.,~\citet{Aumann_IndirectMethodsNuclear_2020,Nunes_Annu.Rev.Nucl.Part.Sci._Nuclear_2020,Shen_ProgressinParticleandNuclearPhysics_Alphacluster_2021,Tumino_Annu.Rev.Nucl.Part.Sci._Trojan_2021a,Schatz_J.Phys.G:Nucl.Part.Phys._Horizons_2022,Liu_NUCLSCITECH_Recent_2024,Tumino_ProgressinParticleandNuclearPhysics_Indirect_2025,Wiescher_Rev.Mod.Phys._Quantum_2025}).

The neutron capture reaction $^{14}$C(n, $\gamma$)$^{15}$C has drawn much attention in the last decades. It was proposed to play a role in several nucleosynthesis processes: (1) it is the slowest reaction in the neutron-induced CNO cycle that takes place in the helium-burning shell of asymptotic giant branch (AGB) stars through the $^{14}$C(n, $\gamma$)$^{15}$C($\beta$-)$^{15}$N(n, $\gamma$)$^{16}$N($\beta$-)$^{16}$O(n, $\gamma$)$^{17}$O(n, $\alpha$)$^{14}$C chain~\citep{Wiescher_BreakoutReactionsCNO_1999}; (2) it is one of the key reactions proposed to be responsible for primordial nucleosynthesis of intermediate-mass elements in the framework of inhomogeneous big-bang nucleosynthesis (IBBN)~\citep{Kajino_PrimordialNucleosynthesisIntermediateMass_1990a,Wiescher_CaptureReactions14C_1990}; (3) it may be present in the main flow paths among the light-mass nuclei in the framework of \textit{r}-process, through the $\alpha$($\alpha$n, $\gamma$)$^{9}$Be($\alpha$, n)$^{12}$C(n, $\gamma$)$^{13}$C(n, $\gamma$)$^{14}$C(n, $\gamma$)$^{15}$C(n, $\gamma$)$\rightarrow$$\cdots$ and the $\alpha$($\alpha$n, $\gamma$)$^{9}$Be(n, $\gamma$)$^{10}$Be($\alpha$, $\gamma$)$^{14}$C(n, $\gamma$)$^{15}$C(n, $\gamma$)$\rightarrow$$\cdots$ chains~\citep{Sasaqui_SensitivityProcessNucleosynthesis_2005b}.

As mentioned above, numerous challenges hinder the precise direct measurement of such (n, $\gamma$) reactions, including the $^{14}$C(n, $\gamma$)$^{15}$C reaction. To date, only three direct measurements of the $^{14}$C(n, $\gamma$)$^{15}$C cross sections have been reported and available~\citep{Beer_Measurement14CGamma_1992, Reifarth_StellarNeutronCapture_2005, Reifarth_14CrossSection_2008}. Among the results of these three measurements, a striking discrepancy by almost a factor of five is revealed. The first direct measurement provided a value of (1.1 $\pm$ 0.28) $\mu$b\footnote{Note that $\mu$b is equal to $\mu$barn where 1 barn $\equiv$ 10$^{-24}$ cm$^{-2}$} \citep{Beer_Measurement14CGamma_1992} at $E_\mathrm{c.m.} = 23.3$ keV\footnote{It should be noted that some articles may exhibit ambiguity or misuse among the cross section at a specific center-of-mass energy $E_\mathrm{c.m.}$, the Maxwellian averaged cross section (MACS) and the spectrum averaged cross section (SACS). In the subsequent discussion of this article, we will provide clear distinctions and primarily focus on the cross section at $E_\mathrm{c.m.} = 23.3$ keV.}, which is approximately five times smaller than the one derived by Wiescher $et\ al$.~\citep{Wiescher_CaptureReactions14C_1990} using the spectroscopic factor deduced by Goss $et\ al$.~\citep{Goss_Phys.Rev.C_Angular_1975}. However, it was found that this experiment was severely influenced by the fact that the container surrounding the $^{14}$C powder was activated by a previous irradiation with an 800 MeV proton beam. Although the latest direct measurement incorporated notable improvements, different analyses of the same dataset still exhibited a threefold discrepancy, primarily due to the dead-time correction~\citep{Reifarth_StellarNeutronCapture_2005, Reifarth_14CrossSection_2008}. However, no further details of the dead-time correction were given.

It is interesting to note that the $^{14}$C(n, $\gamma$)$^{15}$C reaction exhibits certain unique characteristics. 
Firstly, due to the occupancy of the s$_{1/2}$ orbit for the valence neutron, the reaction is dominated by the $p$-wave capture, thus deviating from the 1/$v$ law of the $s$-wave capture and exhibiting a more complex reaction mechanism, as described in Section~\ref{sec:SF_Rate}. 
Secondly, it is one of the rare cases in which the (n, $\gamma$) reaction on unstable nuclei can be studied using both direct and various indirect approaches. So far, apart from the charge symmetry method~\citep{Timofeyuk_15MathrmMathrm_2006}, the primary and common indirect measurement techniques involve the Coulomb dissociation (CD) method and the transfer reaction method. For the Coulomb dissociation method, it typically involves the breakup of $^{15}$C in the Coulomb field of heavy nuclei (e.g. Pb) through time-reversed photodisintegration ($\gamma$,n) to extract the (n, $\gamma$) cross section through the principle of detailed balance~\citep{Horvath_ApJ_Cross_2002,DattaPramanik_CoulombBreakupNeutronrich_2003,Summers_ExtractingEnsuremathGamma_2008,Summers_ErratumExtractingDirect_2008,Nakamura_NeutronCaptureCross_2009,Esbensen_CoulombDissociation15_2009,Esbensen_ErratumCoulombDissociation_2009,Capel_ReconcilingCoulombBreakup_2017,Capel_ErratumReconcilingCoulomb_2018}. For the transfer reaction method, existing studies primarily determine the spectroscopic factor (SF) and asymptotic normalization coefficient (ANC) through the neutron-adding process of $^{14}$C, which, combined with radiative capture theory, is used to deduce the (n, $\gamma$) cross section~\citep{Wiescher_CaptureReactions14C_1990, McCleskey_DeterminationAsymptoticNormalization_2014a, Ma_NewDeterminationAstrophysical_2019}.

\begin{figure}[ht!]
\plotone{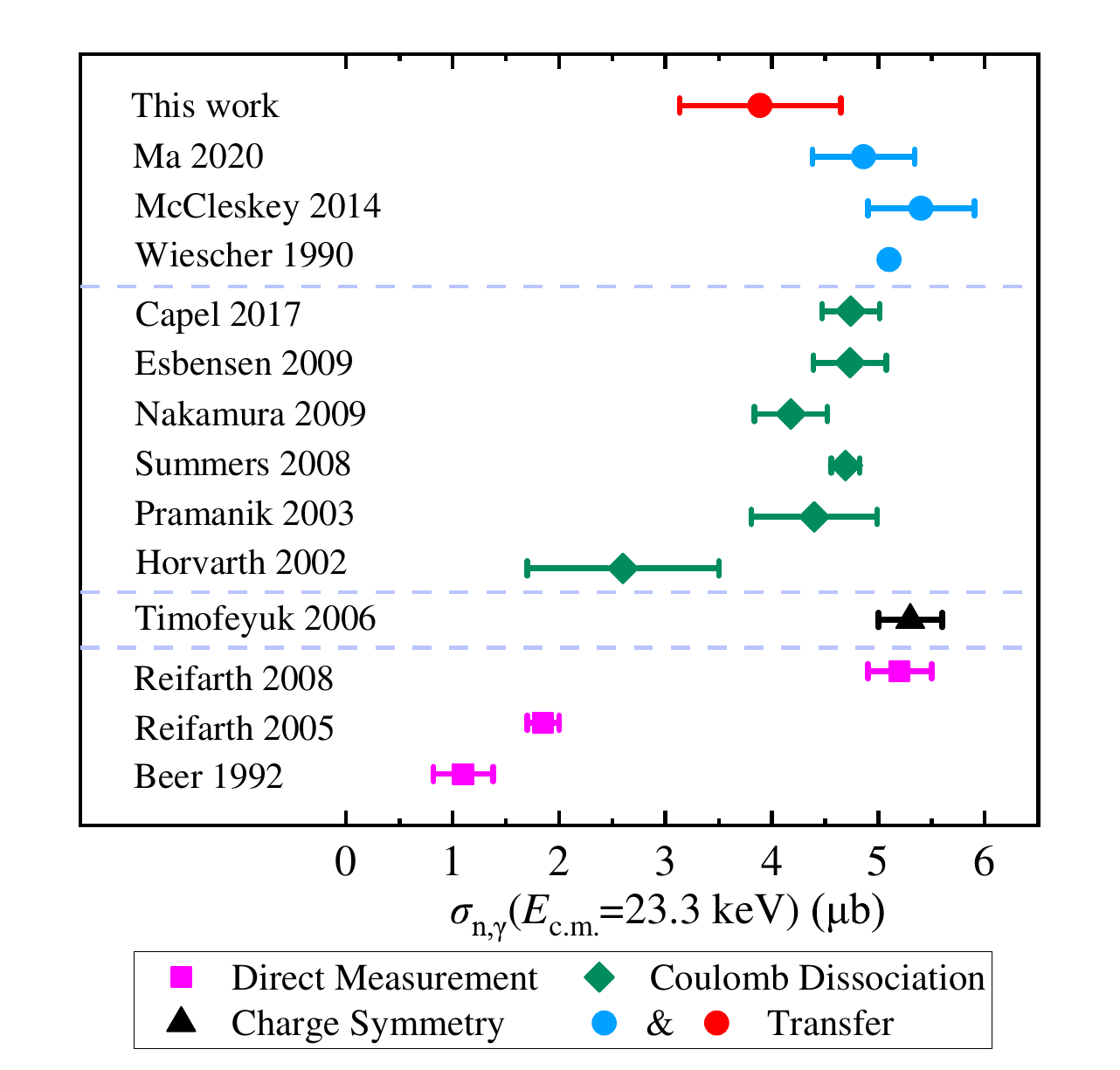}
\caption{Results of the measurement of the $^{14}$C(n, $\gamma$)$^{15}$C reaction cross section, $\sigma(E_\mathrm{_{c.m.}}=23.3\ \mathrm{keV})$, including those using direct measurement (magenta squares), charge symmetry method (black triangles), coulomb dissociation method (green diamonds) and transfer reaction method (red and blue points).}
\label{fig-XSecAt23keV}
\end{figure}

All results of the measurement of $^{14}$C(n, $\gamma$)$^{15}$C reaction cross section, $\sigma(E_\mathrm{_{c.m.}}=23.3\ \mathrm{keV})$, are summarized in Figure~\ref{fig-XSecAt23keV}. Thanks to the availability of multiple complementary indirect approaches, the $^{14}$C(n, $\gamma$)$^{15}$C cross sections can serve as a benchmark to test the validity of modern theoretical frameworks of radiative capture reactions, which in turn is significant for improving our understanding of the stellar evolution and nucleosynthesis. However, it is worth noting that there are some systematic differences between the results from the ANC/SF method (including the transfer reaction method and the charge symmetry method) and those from the Coulomb dissociation method. Furthermore, current transfer reaction studies have certain potential limitations, as all existing work focuses on neutron-adding processes of $^{14}$C (see blue points in Figure~\ref{fig-XSecAt23keV}). Experimental evidence is yet to establish whether there are differences between the neutron- $adding$ and $removing$ processes, while the neutron removal reaction is more similar to the Coulomb dissociation, at least from the perspective of the reaction direction and mechanism.

To clarify the potential differences between the $adding$ and $removing$ processes in transfer reactions and give a more systematic research on the $^{14}$C(n, $\gamma$)$^{15}$C reaction, we have carried out a new experiment using the weakly bound $^{15}$C beam to study single-neutron removal transfer reactions in inverse kinematics. This marks the first study of transfer reactions to infer the cross sections of the $^{14}$C(n, $\gamma$)$^{15}$C reaction from the perspective of the neutron-removing process similar to that in the Coulomb dissociation. In the present experiment, two spectroscopic factors were derived from the (p, d) and (d, t) reaction channels consistently, which were used to deduce the reaction cross section and the corresponding reaction rate. The experimental setup has been shown in detail in~\citet{Jiang_Quenching_2025} and the corresponding data analysis is briefly introduced in Section~\ref{sec:Exp}. The analysis of the new $^{14}$C(n, $\gamma$)$^{15}$C reaction cross section and reaction rate is discussed in Section~\ref{sec:SF_Rate}. Moreover, it is interesting to find that the presented spectroscopic factor is strongly correlated to a long-standing nuclear structure issue so that we are able to simultaneously understand the puzzle of the nuclear structure and the characteristics of this astrophysical reaction. The result sheds light on the impact the microscopic nuclear structure will have on the macroscopic astrophysical applications, as discussed in Section~\ref{sec:NuclStruc}. The implications of the new rate on several astrophysical processes mentioned above are evaluated in Section~\ref{sec:AstroImpli} using state-of-the-art theoretical models to re-investigate the impact of this reaction within a modern framework. Finally, we summarize the present study in Section~\ref{sec:conclusion}.


\section{The Experiment of Single-Neutron Removal Transfer Reactions on $^{15}$C}~\label{sec:Exp}

The experiment was carried out using the HELIOS spectrometer~\citep{Lighthall_NuclearInstrumentsandMethodsinPhysicsResearchSectionA:AcceleratorsSpectrometersDetectorsandAssociatedEquipment_Commissioning_2010} at the ATLAS facility at the Argonne National Laboratory. A 7.1 MeV/u $^{15}$C secondary beam was produced at the ATLAS in-flight system, RAISOR~\citep{Hoffman_NuclearInstrumentsandMethodsinPhysicsResearchSectionA:AcceleratorsSpectrometersDetectorsandAssociatedEquipment_Flight_2022} 
with an intensity of approximately 10$^{6}$ particles per second with negligible contamination (less than 1\%). The experimental setup and procedures are similar to those previously reported~\citep{Hoffman_Phys.Rev.C_Experimental_2012a, Kay_Phys.Rev.Lett._Quenching_2022} and has already been elaborated in detail in another article~\citep{Jiang_Quenching_2025}.
The $^{15}$C beam bombarded a target of deuterated polyethylene (CD$_{2}$)$_{n}$ of thickness 363(20) $\mu$g/cm$^{2}$ or polyethylene (CH$_{2}$)$_{n}$ of thickness 387(22) $\mu$g/cm$^{2}$, which corresponded to the measurement of $^{15}$C(p, d)$^{14}$C and $^{15}$C(d, t)$^{14}$C, respectively. 

\begin{figure}[ht!]
\plotone{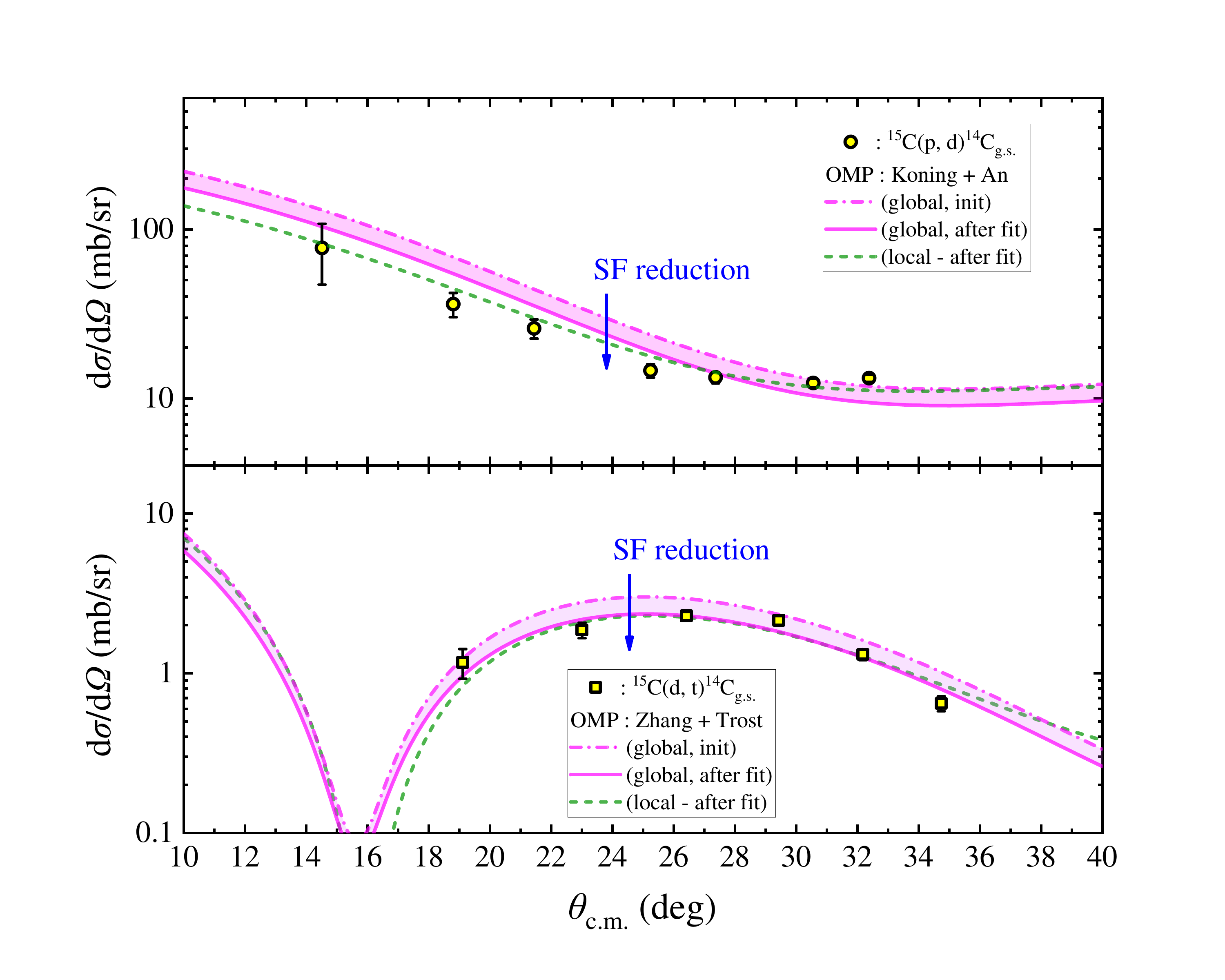}
\caption{The differential cross sections for $^{15}$C(p, d)$^{14}$C$\mathrm{_{g.s.}}$ (yellow circles, top panel) and $^{15}$C(d, t)$^{14}$C$\mathrm{_{g.s.}}$ (yellow squares, bottom panel), respectively. The dashed-dotted magenta lines represent typical DWBA calculations using global OMPs for entrance and exit channels. Results of parameterizations from~\citet{Koning_LocalGlobalNucleon_2003,An_GlobalDeuteronOptical_2006} for (p, d) channel and those of parameterizations from~\citet{Zhang_OpticalModelPotential_2016,Trost_SimpleOpticalModel_1987} for (d, t) channel are shown, respectively. The solid lines represent the theoretical differential cross sections which have been normalized to the experimental data and are calculated using global OMPs, while the dashed lines represent the counterparts using local OMPs derived from fitting the elastic scattering data. The magenta bands and blue arrows are to guide the eye and illustrate the so-called ``quenching'' effect in the nuclear structure field (see text). Note that as the theoretical SF of $^{15}$C (0.95) calculated by the independent particle shell model is almost near unity, the experimental SF is almost equal to the quenching factor.
\label{fig-AngularDistribution}}
\end{figure}

The particle identification (PID) was performed by analyzing the cyclotron period of emitted light charged particles in a uniform magnetic field, combined with four sets of $\Delta$$E$-$E$ telescopes to detect $^{14}$C residues. The intensity of the incoming beam was monitored using a downstream ionization chamber. The kinematic reconstruction was achieved through high-precision position-sensitive silicon detector (PSD) arrays, which provided both position and energy information. This approach enabled the extraction of differential cross sections for transfer reactions populating the ground state (g.s.) of $^{14}$C (shown in Figure~\ref{fig-AngularDistribution}), covering a range of center-of-mass angles ($\theta_\mathrm{C.M.}$) from 15$^{\circ}$ to 35$^{\circ}$. 
The uncertainties of the differential cross section mainly come from the statistical error and the systematic uncertainties resulting from the acceptance correction and the timing gate efficiency correction (see~\citet{Jiang_Quenching_2025} for details). In the following section, we will go over how to use distorted-wave Born approximation (DWBA) theory to extract SF/ANC of the $^{15}$C g.s., thereby deriving the reaction cross section and reaction rate of the $^{14}$C(n, $\gamma$)$^{15}$C reaction.


\section{Extracting the SF/ANC of $^{15}$C and Stellar Reaction Rate of $^{14}$C(\MakeLowercase{n}, $\gamma$)$^{15}$C} \label{sec:SF_Rate}

In this energy regime, 
such reactions are extremely fast, leading to fewer internal collisions. This kind of reaction typically exhibits high cross sections at forward angles. The finite-range DWBA theory, which assumes a one-step transition between the initial and final scattering states, is commonly used to interpret direct reaction mechanisms.

In the $^{15}$C(p, d)$^{14}$C$\mathrm{_{g.s.}}$ and $^{15}$C(d, t)$^{14}$C$\mathrm{_{g.s.}}$ reactions, the experimental differential cross section can be related to the theoretical DWBA calculation by\footnote{Note that for the (p, d) and (d, t) channel calculations, the neutron SFs in deuterium and tritium have been incorporated in the bound state calculations of~\citet{Wiringa_AccurateNucleonnucleonPotential_1995} and~\citet{Brida_Phys.Rev.C_Quantum_2011}. Therefore, we will obtain the simplified equation~\ref{DWBA} which omitts neutron SFs in deuterium and tritium instead of the complete form shown in literature such as~\citet{Mukhamedzhanov_Phys.Rev.C_Asymptotic_2001,Tribble_RepProgPhys_IndirectTechniquesNuclear_2014}.}
\begin{equation}
\label{DWBA}
(\frac{d \sigma}{d \Omega})\mathrm{_{exp}} = \mathrm{SF_{^{15}C, exp}} \cdot (\frac{d \sigma}{d \Omega})\mathrm{_{DWBA}},
\end{equation}
where $(\frac{d \sigma}{d \Omega})\mathrm{_{exp}}$ and $(\frac{d \sigma}{d \Omega})\mathrm{_{DWBA}}$ denote the experimental differential cross section and that derived by the conventional DWBA calculation, respectively. $\mathrm{SF_{^{15}C, exp}}$ is the experimental spectroscopic factor, which in principle reflects the occupancy of a certain orbit. In addition, by defining the amplitude of the tail of the radial overlap function, we obtain another normalization equation
\begin{equation}
\label{DWBA_ANC}
(\frac{d \sigma}{d \Omega})\mathrm{_{exp}} = \frac{(C\mathrm{_{n+^{14}C}^{^{15}C_{g.s.}}})^2}{(b\mathrm{_{n+^{14}C}^{^{15}C_{g.s.}}})^2} \cdot (\frac{d \sigma}{d \Omega})\mathrm{_{DWBA}},
\end{equation}
where $C\mathrm{_{n+^{14}C}^{^{15}C_{g.s.}}}$ is the ANC for the valence neutron bound state in the $^{15}$C and $b\mathrm{_{n+^{14}C}^{^{15}C_{g.s.}}}$ is the corresponding single-particle ANC. The single-particle ANC be derived by analyzing the asymptotic property of the single-particle wave function, as shown in
\begin{equation}
\label{single_ANC}
\phi(r) \approx  b\mathrm{_{n+^{14}C}^{^{15}C_{g.s.}}} \cdot \frac{W_{-\eta, l+1/2} (2kr)}{r},
\end{equation}

In equation~\ref{single_ANC}, $\phi(r)$ is radial bound state wave function of $^{15}$C, which is calculated using the single-particle potential. $W_{-\eta, l+1/2} (2kr)$ is the Whittaker function for the bound state, where $\eta$ is the Sommerfeld parameter for the bound state ($\eta$=0 in this case because of the lack of charge in neutrons). $l$ is the relative orbital angular momentum of the valence neutron, $k$ is the bound state wave number, and $r$ is the radial parameter in the center-of-mass frame. By comparing the asymptotic form of the single-particle wave function with the Whittaker function, a single-particle ANC can be obtained. Finally, combining equation~\ref{DWBA} and~\ref{DWBA_ANC}, the ANC can be related to SF by equation~\ref{SF_ANC}. Therefore, to extract the ANC or SF of $^{15}$C g.s., the DWBA calculations are normalized to the experimental data
\begin{equation}
\label{SF_ANC}
(C\mathrm{_{n+^{14}C}^{^{15}C_{g.s.}}})^2 = (b\mathrm{_{n+^{14}C}^{^{15}C_{g.s.}}})^2 \cdot \mathrm{SF_{^{15}C, exp}}
\end{equation}

DWBA calculations were made using the finite range code PTOLEMY~\citep{Macfarlane_PTOLEMYProgramHeavyIon_1978}. The essential ingredients of these calculations are the optical model potentials (OMPs) for the entrance and exit channels and the potentials for the bound states of the compound nuclei. The complete parameter set used in the DWBA calculations is beyond the scope of this paper and can be found in~\citet{Jiang_Quenching_2025}.

The final results of the SFs are shown in Table~\ref{tab:SFAndUncertainty}, with uncertainties coming mainly from the target thickness ($\approx$ 6\%), the beam intensity ($\approx$ 15\%) and the SF fitting using local and global OMP parameter sets ($\approx$22\% for (p, d) and $\approx$24\% for (d, t)). $\mathrm{SF_{^{15}C, exp}}$= 0.68 $\pm$ 0.14 was determined by averaging the results of two reaction channels, since they are probing the same spectroscopic factor and we assume that the results obtained from different reaction channels but with the same reaction direction should be consistent. Our new result is consistent with the previous work on $^{15}$C using (d, p) neutron-$adding$ channel~\citep{Kay_Phys.Rev.Lett._Quenching_2022}. It is also interesting to find that, as reviewed in ~\citet{Mukhamedzhanov_AsymptoticNormalizationCoefficients_2011a}, there is an overestimation of a factor of 1.293 for the absolute value of the differential cross section derived in~\citet{Goss_Phys.Rev.C_Angular_1975}. We note that if the SF deduced in~\citet{Goss_Phys.Rev.C_Angular_1975} (0.88) is divided by a factor of 1.293, the resulting SF (0.68) aligns perfectly with our result.

\begin{deluxetable*}{cccc}
\tablecaption{List of SF Deduced From Different Reaction Channels and the Corresponding Main Sources of Uncertainties}
\label{tab:SFAndUncertainty}
\tablehead
{
\colhead{Reaction Channel} & \colhead{Parameter} & \colhead{$1\sigma$ Uncertainty} & \colhead{SF(exp)}
}
\startdata
\multirow{3}{*}{$^{15}$C(p, d)$^{14}$C$_{\mathrm{g.s.}}$} & Target thickness & 6\% & \multirow{3}{*}{0.67(18)} \\
 & Beam intensity & 15\% & \\
 & SF fitting & 22\% & \\
\multirow{3}{*}{$^{15}$C(d, t)$^{14}$C$_{\mathrm{g.s.}}$} & Target thickness & 6\% & \multirow{3}{*}{0.70(20)} \\
 & Beam intensity & 15\% & \\
 & SF fitting & 24\% & \\
\hline
Total uncertainty after average & & 21\% & 0.68(14) \\
\enddata
\end{deluxetable*}

By calculating the radial bound state wave function using the bound state parameters of $^{15}$C mentioned above and comparing it to the Whittaker function in the asymptotic region, we determined the single-particle ANC $b\mathrm{_{n+^{14}C}^{^{15}C_{g.s.}}}$ to be 1.41 fm$^{-1/2}$. Therefore, the ANC$^{2}$ of $^{15}$C$\mathrm{_{g.s.}}$ ($(C\mathrm{_{n+^{14}C}^{^{15}C_{g.s.}}})^2$) was deduced to be (1.35 $\pm$ 0.26) fm$^{-1}$. Although the bound state parameters affect the values of the single-particle ANC and SF, we observed that a 7\% increase(decrease) in the SF corresponded to a decrease(increase) in the ANC of less than 4\%. Therefore, we conclude that the uncertainties of ANC from bound state parameters are very limited.

The comparison of SF / ANC derived from different methods is listed in Table \ref{tab:sf_anc}. Our new ANC result (1.35$\pm$0.26) fm$^{-1}$ is the smallest ever determined in the transfer framework. It is consistent with the value derived in~\citet{Pang_AsymptoticNormalizationCoefficients_2014,Mukhamedzhanov_AsymptoticNormalizationCoefficients_2011a}, but contradicts those derived in~\citet{McCleskey_DeterminationAsymptoticNormalization_2014a, Summers_ErratumExtractingDirect_2008, Timofeyuk_15MathrmMathrm_2006}.

\begin{deluxetable*}{lccccc}
\tablecaption{The SFs and ANCs of $^{15}$C$\mathrm{g.s.}$ from different methods. CD represents the Coulomb dissociation. PWA, DWBA, ADWA and CDCC represent the plane-wave approximation, distorted wave Born approximation, adiabatic distorted wave approximation, and continuum discretized coupled channel analysis, respectively. 
\label{tab:sf_anc}}
\tablehead{
\colhead{Reference} & 
\colhead{Experiment or Data Set} & 
\colhead{Method} & 
\colhead{SF} & 
\colhead{ANC$^2$ (fm$^{-1}$)}
}
\startdata
\citet{DattaPramanik_CoulombBreakupNeutronrich_2003} & CD of $^{15}$C \citep{DattaPramanik_CoulombBreakupNeutronrich_2003} & PWA & 0.73 $\pm$ 0.05\tablenotemark{a} & $-$ \\
 &  & DWBA & 0.97 $\pm$ 0.08\tablenotemark{a} & $-$ \\
\citet{Summers_ExtractingEnsuremathGamma_2008, Summers_ErratumExtractingDirect_2008} & CD of $^{15}$C \citep{Nakamura_NuclearPhysicsA_Coulomb_2003} & CDCC & $-$ & 1.74 $\pm$ 0.11 \\
\citet{Nakamura_NeutronCaptureCross_2009} & CD of $^{15}$C \citep{Nakamura_NeutronCaptureCross_2009} & $-$ & 0.91 $\pm$ 0.06\tablenotemark{b} & $-$ \\
\citet{Timofeyuk_15MathrmMathrm_2006} & $^{15}$F width \citep{Timofeyuk_15MathrmMathrm_2006} & charge symmetry & $-$ & $1.89 \pm 0.11$ \\
\citet{Goss_Phys.Rev.C_Angular_1975} & $^{14}$C(d, p)$^{15}$C \citep{Goss_Phys.Rev.C_Angular_1975} & DWBA & 0.88 & 1.96 \\
\citet{Pang_AreSpectroscopicFactors_2007} & $^{14}$C(d, p)$^{15}$C \citep{Goss_Phys.Rev.C_Angular_1975} & DWBA & 1.27\tablenotemark{c} & 2.70\tablenotemark{c} \\
 & $^{14}$C(d, p)$^{15}$C \citep{Goss_Phys.Rev.C_Angular_1975} & ADWA & 1.01\tablenotemark{c} & 2.14\tablenotemark{c} \\
\citet{Mukhamedzhanov_AsymptoticNormalizationCoefficients_2011a} & $^{14}$C(d, p)$^{15}$C \citep{Mukhamedzhanov_AsymptoticNormalizationCoefficients_2011a} & ADWA & $-$ & 1.64 $\pm$ 0.26 \\
 & & DWBA & $-$ & 1.92 $\pm$ 0.46 & \\
\citet{McCleskey_DeterminationAsymptoticNormalization_2014a} & $^{13}$C($^{14}$C,$^{15}$C)$^{12}$C \citep{McCleskey_DeterminationAsymptoticNormalization_2014a} & DWBA & 1.12 & 2.09 $\pm$ 0.29 & \\
 & d($^{14}$C, p)$^{15}$C \citep{McCleskey_DeterminationAsymptoticNormalization_2014a} & ADWA & $-$ & 1.77 $\pm$ 0.21 \\
 & & Average & $-$ & 1.88 $\pm$ 0.18 \\
\citet{Pang_AsymptoticNormalizationCoefficients_2014} & $^{14}$C(d, p)$^{15}$C \citep{McCleskey_DeterminationAsymptoticNormalization_2014a} & CDCC & 0.82 $\pm$ 0.03 & 1.80 $\pm$ 0.2 \\
\citet{Kay_Phys.Rev.Lett._Quenching_2022} & d($^{14}$C, p)$^{15}$C \citep{Kay_Phys.Rev.Lett._Quenching_2022} & DWBA\&ADWA & 0.51 $\pm$ 0.12 & $-$ \\
This work & p($^{15}$C, d)$^{14}$C \& d($^{15}$C, t)$^{14}$C & DWBA & 0.68 $\pm$ 0.14 & 1.35 $\pm$ 0.26 \\
\enddata
\tablenotetext{a}{derived with $r_{\mathrm{0}}$ = 1.25 fm, $a_{\mathrm{0}}$ = 0.7 fm, see  \citet{DattaPramanik_CoulombBreakupNeutronrich_2003}.}
\tablenotetext{b}{derived with $r_{\mathrm{0}}$ = 1.223 fm, $a_{\mathrm{0}}$ = 0.5 fm, see Tabel I in  \citet{Nakamura_NeutronCaptureCross_2009}.}
\tablenotetext{c}{derived with $r_{\mathrm{0}}$ = 1.31 fm, $b$ = 1.45 fm$^{-1}$, see Table III in  \citet{Pang_AreSpectroscopicFactors_2007}.}
\end{deluxetable*}

\begin{figure}[ht!]
\plotone{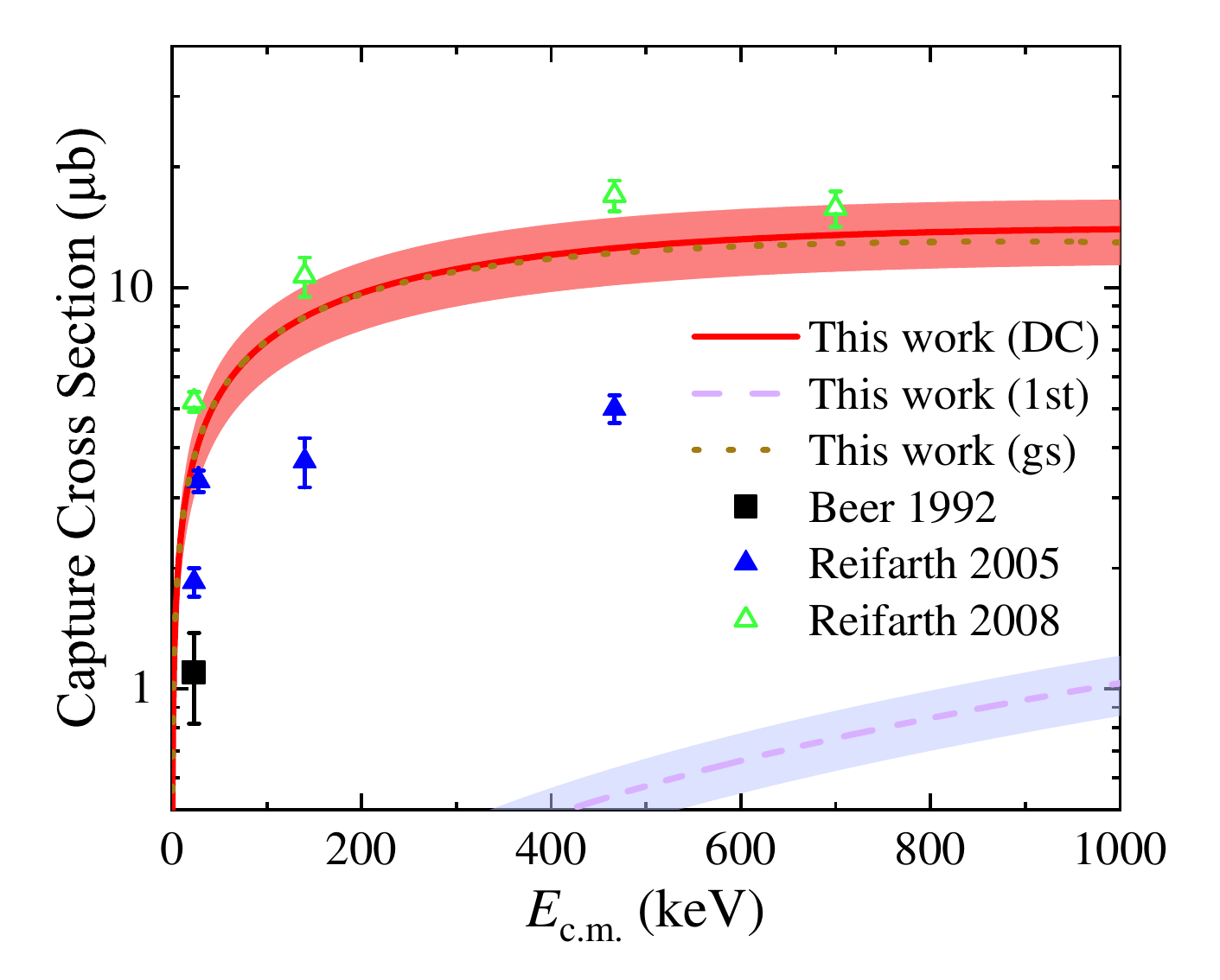}
\caption{The DC cross sections of the $^{14}$C(n, $\gamma$)$^{15}$C reaction. 
The red line and band (labeled ``DC'') show the total DC cross sections. The dark-yellow line (labeled ``gs'') shows the cross sections of DC to the ground state of $^{15}$C using the SF and the bound states parameters determined in this work. The purple line and band (labeled ``1st'') show the cross sections of DC to the 1$^{\text{st}}$ excited state of $^{15}$C using the SF and the bound states parameters determined in~\citet{Kay_Phys.Rev.Lett._Quenching_2022}. Note that since the cross sections of DC to the 1$^{\text{st}}$ excited state are much smaller than those of DC to the ground state, we do not show the dark-yellow band to avoid visual overlap with the red band.}
\label{fig-TotXsec}
\end{figure}

With the parameters determined above, we used the code RADCAP~\citep{Bertulani_ComputerPhysicsCommunications_RADCAP_2003} to calculate the radiative capture cross sections of the $^{14}$C(n, $\gamma$)$^{15}$C reaction. Due to the simple fact that the neutron separation energy of $^{15}$C is only 1.218 MeV, there are no low-lying resonances in astrophysical environments. Therefore, the $E\mathrm{_{1}}$ transition from the incoming $p$-wave to the 2$s\mathrm{_{1/2}}$ ground state will dominate, with the energy dependence shown in equation~\ref{Eq:pwave-capture}, where $E_{\mathrm{B}}$, $\mu_{M}$, $\hbar$, $c$, $e$, $Z$, $A$ denote the binding energy of the bound state, the reduced mass of the particle pair (here refering to n+$^{14}$C), the reduced Planck constant, the speed of light, the elementary charge, the number of protons of the nucleus, and the mass number of the nucleus~\citep{Bertulani_ComputerPhysicsCommunications_RADCAP_2003,Otsuka_Phys.Rev.C_Neutron_1994}. The SF of the first excited state ($E\mathrm{_{x}}$ = 0.74 MeV) was taken to be 0.41(7) from ~\citet{Kay_Phys.Rev.Lett._Quenching_2022}. As suggested in ~\citet{Huang_RadiativeCaptureNucleons_2010}, the bound state parameters need to be the same as those used when extracting the SF. Therefore, with the parameters defined above, direct capture (DC) cross sections were derived. The total DC cross sections of the $^{14}$C(n, $\gamma$)$^{15}$C reaction are shown in Figure~\ref{fig-TotXsec}. Considering that the resonance capture to the second excited state is too weak below 1 MeV, we give $\sigma(E_\mathrm{_{c.m.}}=23.3\ \mathrm{keV})$ = 3.89 $\pm$ 0.76 $\mu$b (see the red point in Figure~\ref{fig-XSecAt23keV}), which shows more consistency with previous Coulomb dissociation work, while it is approximately 30\% smaller than that deduced from the latest direct measurement~\citep{Reifarth_14CrossSection_2008}.

\begin{equation}
\label{Eq:pwave-capture}
\sigma_{E_{\mathrm{1}}}^{\text{non-resonance}}(p-\text{wave}) \propto \frac{16}{3}\frac{\hbar c}{(\mu_{M}c^{2})^2}e^{2}(\frac{Z}{A})^{2}\frac{\sqrt{EE_{B}}}{E + E_{B}} 
\end{equation}

\begin{figure}[ht!]
\epsscale{0.7}
\plotone{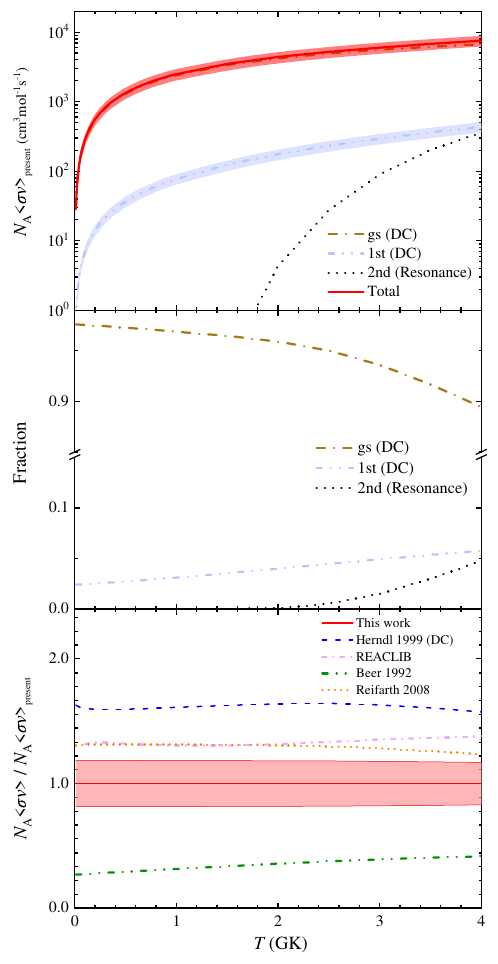}
\caption{The $^{14}$C(n, $\gamma$)$^{15}$C reaction rate of the present work. Top panel: The red solid line and band are the total reaction rate and the corresponding 1$\sigma$ uncertainty, respectively. The dashed-dotted dark-yellow, dash-double-dotted purple, dotted black lines represent the contributions from DC to the ground state (labeled ``gs (DC)"), DC to the first excited state (labeled ``1st (DC)"), and the resonance capture to the second excited state (labeled ``2nd (Resonance)"), respectively. Note that here we do not show the uncertainty band of DC to the ground state with the same reason declared in Figure~\ref{fig-TotXsec}; Middle panel: The average fractions of the DC and resonance components, respectively, with the same colors in the top panel; Bottom panel: The ratios of previous reaction rates to our new result (solid red line and the shaded area). The previous reaction rates were adapted from~\citet{Herndl_ReactionRatesNeutron_1999a} (dashed blue line),~\citet{Cyburt_JINAREACLIBDATABASE_2010} (dashed-dotted pink line),~\citet{Rauscher_ProductionHeavyElements_1994} (dashed-double-dotted green line), and~\citet{Reifarth_14CrossSection_2008} (dotted orange line).}
\label{fig-ReactionRate}
\end{figure}

The reaction rate was calculated by 
\begin{equation}
\label{ReactionRate}
N_{A}\langle\sigma v\rangle = N_{A}(\frac{8}{\pi \mu_{M}})^{1/2} (kT)^{-3/2}\int_{0}^{\infty}E_{\mathrm{c.m.}}\sigma(E_{\mathrm{c.m.}})e^{-E_{\mathrm{c.m.}}/kT}dE_{\mathrm{c.m.}},
\end{equation}
where $N_{A}$, $k$, $T$ are the Avogadro constant, the Boltzmann’s constant and the stellar temperature, respectively. Furthermore, the contribution of resonant capture to the second excited state ($E\mathrm{_{x}}$ = 3.10 MeV) was taken into account using the Breit-Wigner resonant equation, while a neutron width of $\Gamma\mathrm{_{n}}$ = 42 keV and a gamma width of $\Gamma\mathrm{_{\gamma}}$ = 4 eV was chosen \citep{Wiescher_CaptureReactions14C_1990}. For such a narrow resonance, the contribution to the total reaction rate is approximately given by 
\begin{equation}
\label{Eq-ReactionRate}
N_{A}\langle\sigma v\rangle = N_{A}(\frac{2\pi}{\mu_{M} kT})^{3/2} \hbar^{2}\omega\gamma \exp(-\frac{E_{\mathrm{R}}}{kT}),
\end{equation}
where $\hbar$, $\omega \gamma$, $E_{\mathrm{R}}$ are the reduced Planck constant, resonance strength and the resonance energy, respectively. The resonance strength in equation~\ref{Eq-ReactionRate} was deduced from the neutron and gamma widths above. The total reaction rate, fractions of partial reaction rate, and the comparison with previous rates between 0.01-4 GK are shown in Figure~\ref{fig-ReactionRate}. It can be seen that our rate is approximately 25\% lower than that derived in ~\citet{Reifarth_14CrossSection_2008}, and approximately 2.5-4 times higher than that derived in ~\citet{Beer_Measurement14CGamma_1992}. The reaction rate was fitted to the standard parametrization adopted by REACLIB~\citep{Cyburt_JINAREACLIBDATABASE_2010}
\begin{equation}
\label{ReactionRate-Fit}
N_{A}\langle\sigma v\rangle = \mathrm{exp}(a_{1} + a_{2}T_{9}^{-1} + a_{3}T_{9}^{-1/3} + a_{4}T_{9}^{1/3} + a_{5}T_{9} + a_{6}T_{9}^{5/3} + a_{7}\mathrm{ln}(T_{9})),
\end{equation}
where the reaction rate is given in cm$^{3}$s$^{-1}$mol$^{-1}$ and $T_{9}$ is the stellar temperature in 10$^{9}$ K. The total reaction rate of $^{14}$C(n, $\gamma$)$^{15}$C is tabulated in Table \ref{tab:reactionrate_compare} and the best-fit parameters which reproduce the numerical values to within 1\% in the range of 0.01 $\leq$ $T_{9}$ $\leq$ 10 are given by
\begin{equation}
\label{ReactionRate-FitPara}
\begin{split}
a_{1} &= 1.051 \times 10^{1}, \\
a_{2} &= -6.573 \times 10^{-3}, \\
a_{3} &= 9.771 \times 10^{-1}, \\
a_{4} &= -3.967 \times 10^{0}, \\
a_{5} &= 3.226 \times 10^{-1}, \\
a_{6} &= -2.485 \times 10^{-2}, \\
a_{7} &= 2.209 \times 10^{0}. \\
\end{split}
\end{equation}

\begin{deluxetable*}{cccccc}
\tablecaption{The reaction rate determined in this work and comparison with previous results.}
\label{tab:reactionrate_compare}
\tablehead
{
\colhead{$T_9$} & 
\multicolumn{2}{c}{Present reaction rate (cm$^3$ s$^{-1}$ mol$^{-1}$)} & & \multicolumn{2}{c}{Ratio\tablenotemark{a}} \\
\cline{2-3} \cline{5-6}
\colhead{} & \colhead{Central value} & \colhead{1$\sigma$ uncertainty} & &\colhead{Beer1992} & \colhead{Reifarth2008}
}
\startdata
0.01 & 2.89 $\times$ 10$^{1}$ & 5.37 $\times$ 10$^{0}$ & & 3.71 & 0.76 \\
0.02 & 5.79 $\times$ 10$^{1}$ & 1.08 $\times$ 10$^{1}$ & & 3.71 & 0.76 \\
0.03 & 8.67 $\times$ 10$^{1}$ & 1.61 $\times$ 10$^{1}$ & & 3.70 & 0.76 \\
0.04 & 1.15 $\times$ 10$^{2}$ & 2.14 $\times$ 10$^{1}$ & & 3.70 & 0.76 \\
0.05 & 1.44 $\times$ 10$^{2}$ & 2.67 $\times$ 10$^{1}$ & & 3.69 & 0.76 \\
0.06 & 1.73 $\times$ 10$^{2}$ & 3.20 $\times$ 10$^{1}$ & & 3.68 & 0.76 \\
0.07 & 2.01 $\times$ 10$^{2}$ & 3.73 $\times$ 10$^{1}$ & & 3.68 & 0.76 \\
0.08 & 2.29 $\times$ 10$^{2}$ & 4.26 $\times$ 10$^{1}$ & & 3.67 & 0.75 \\
0.09 & 2.57 $\times$ 10$^{2}$ & 4.78 $\times$ 10$^{1}$ & & 3.67 & 0.75 \\
0.1 & 2.86 $\times$ 10$^{2}$ & 5.30 $\times$ 10$^{1}$ & & 3.66 & 0.75 \\
0.15 & 4.25 $\times$ 10$^{2}$ & 7.88 $\times$ 10$^{1}$ & & 3.63 & 0.75 \\
0.2 & 5.61 $\times$ 10$^{2}$ & 1.04 $\times$ 10$^{2}$ & & 3.60 & 0.75 \\
0.25 & 6.96 $\times$ 10$^{2}$ & 1.29 $\times$ 10$^{2}$ & & 3.57 & 0.75 \\
0.3 & 8.28 $\times$ 10$^{2}$ & 1.54 $\times$ 10$^{2}$ & & 3.54 & 0.75 \\
0.35 & 9.58 $\times$ 10$^{2}$ & 1.78 $\times$ 10$^{2}$ & & 3.51 & 0.75 \\
0.4 & 1.09 $\times$ 10$^{3}$ & 2.01 $\times$ 10$^{2}$ & & 3.48 & 0.75 \\
0.45 & 1.21 $\times$ 10$^{3}$ & 2.25 $\times$ 10$^{2}$ & & 3.45 & 0.75 \\
0.5 & 1.34 $\times$ 10$^{3}$ & 2.48 $\times$ 10$^{2}$ & & 3.43 & 0.75 \\
0.55 & 1.46 $\times$ 10$^{3}$ & 2.70 $\times$ 10$^{2}$ & & 3.40 & 0.75 \\
0.6 & 1.58 $\times$ 10$^{3}$ & 2.92 $\times$ 10$^{2}$ & & 3.38 & 0.75 \\
0.65 & 1.70 $\times$ 10$^{3}$ & 3.14 $\times$ 10$^{2}$ & & 3.35 & 0.75 \\
0.7 & 1.82 $\times$ 10$^{3}$ & 3.36 $\times$ 10$^{2}$ & & 3.33 & 0.75 \\
0.75 & 1.93 $\times$ 10$^{3}$ & 3.57 $\times$ 10$^{2}$ & & 3.30 & 0.75 \\
0.8 & 2.04 $\times$ 10$^{3}$ & 3.78 $\times$ 10$^{2}$ & & 3.28 & 0.75 \\
0.85 & 2.16 $\times$ 10$^{3}$ & 3.98 $\times$ 10$^{2}$ & & 3.25 & 0.75 \\
0.9 & 2.27 $\times$ 10$^{3}$ & 4.19 $\times$ 10$^{2}$ & & 3.23 & 0.75 \\
0.95 & 2.38 $\times$ 10$^{3}$ & 4.39 $\times$ 10$^{2}$ & & 3.21 & 0.75 \\
1 & 2.48 $\times$ 10$^{3}$ & 4.58 $\times$ 10$^{2}$ & & 3.18 & 0.75 \\
2 & 4.38 $\times$ 10$^{3}$ & 8.00 $\times$ 10$^{2}$ & & 2.81 & 0.76 \\
2.5 & 5.20 $\times$ 10$^{3}$ & 9.41 $\times$ 10$^{2}$ & & 2.66 & 0.76 \\
3 & 5.98 $\times$ 10$^{3}$ & 1.07 $\times$ 10$^{3}$ & & 2.56 & 0.78 \\
3.5 & 6.74 $\times$ 10$^{3}$ & 1.18 $\times$ 10$^{3}$ & & 2.47 & 0.79 \\
4 & 7.50 $\times$ 10$^{3}$ & 1.28 $\times$ 10$^{3}$ & & 2.40 & 0.80 \\
\enddata
\tablenotemark{a}{Here it is the ratio of our reaction rate to previous rates.}
\tablecomments{We list the reaction rates within the temperature range of 0.01 $\leq$ $T_{9}$ $\leq$ 4 for the convenience of comparison, since the reaction rate in ~\citet{Reifarth_14CrossSection_2008} were provided only up to a maximum temperature of 4 GK.}
\end{deluxetable*}

\section{The Correlation With Nuclear Structure Issues} \label{sec:NuclStruc}
The independent particle model based on mean-field theory serves as a fundamental cornerstone of modern nuclear structure research. Over the past two decades, experimental observations from transfer reactions and electron-induced knockout (e, e'p) reactions have revealed a significant reduction of single-particle strength (i.e., SF) within the nuclei, with the observed strength being only approximately 60\% of the predictions made by the independent particle model~\citep{Kramer_ConsistentAnalysis3He_2001}. This ratio of experimentally measured SF to that deduced from theoretical predictions is commonly referred to as the ``quenching" factor $R_{\mathrm{s}}$. Clearly, the quenching effect and its systematic behavior are of critical importance. This is because the quenching effect determines the magnitude of the SF, which the direct capture reaction cross section is proportional to. Consequently, the quenching effect directly influences the magnitudes of nuclear reaction cross sections and the corresponding reaction rates, thus playing a decisive role in shaping the outcomes of stellar evolution. 

Recent systematic studies of heavy-ion (HI) induced single-nucleon knockout reactions uncovered a surprising and puzzling phenomenon, that $R_{\mathrm{s}}$ exhibits a strong negative dependence on the Fermi surface asymmetry $\Delta S$, where \rm{$\Delta$}$S$ is defined as $S\mathrm{_{p}}$-$S\mathrm{_{n}}$/$S\mathrm{_{n}}$-$S\mathrm{_{p}}$ for proton/neutron removal to the g.s., respectively~\citep{Gade_ReductionSpectroscopicStrength_2008, Tostevin_SystematicsIntermediateenergySinglenucleon_2014, Tostevin_UpdatedSystematicsIntermediateenergy_2021}. However, this dependence has not been observed in experiments involving transfer reactions, electron-induced knockout (e, e'p) reactions, or proton-induced knockout (p, 2p) reactions (see, e.g.,~\citet{Kramer_ConsistentAnalysis3He_2001, Manfredi_Phys.Rev.C_Quenching_2021, R^3BCollaboration_Quasifree2pReactions_2018}). It is evident that different systematics of the ``quenching'' effect will greatly change our understanding of numerous nuclear astrophysical reactions involving various isotopes within a wide range of $\Delta S$.

Regrettably, due to the limited experimental techniques available for studying unstable nuclei in the past, there is currently a scarcity of data in the framework of transfer reactions for regions with $\Delta S$ $\leq$ -15 MeV, which in turn casts a shadow over our accurate understanding of nucleosynthesis in astrophysical environments. The only two recent experiments that have explored the quenching puzzle in such exotic systems are the studies of the $^{15}$C nuclei~\citep{Kay_Phys.Rev.Lett._Quenching_2022} and the $^{14}$O nuclei~\citep{Flavigny_LimitedAsymmetryDependence_2013a}. As discussed above, no experimental evidence has yet confirmed whether there are potential differences between neutron $adding$ and $removing$ processes. It also remains a puzzle whether there is a strong dependence for transfer reactions just like that observed in HI-induced knockout. This work, however, investigated the neutron $removal$ reactions for the first time in the transfer framework and obtained consistent experimental SFs across two different reaction channels (see Figure~\ref{fig-AngularDistribution}). Furthermore, our result is in agreement within uncertainties with that derived from the neutron $adding$ process, which will play a significant role in the advance of the study and compilation of the systematics of transfer reactions and the related nuclear astrophysical reactions~\citep{Lee_NeutronSpectroscopicFactors_2007}. Therefore, we conclude that reliable SFs have been obtained in the most extreme $\Delta S$ region achieved so far for transfer reactions, supporting the hypothesis that the strong negative dependence exists only in HI-induced knockout reactions. Our work acts as a bridge between nuclear structure and nuclear astrophysics, with the astrophysical implications of our new result discussed in the next sections.

\section{Astrophysical Implications} \label{sec:AstroImpli}

\subsection{Nucleosynthesis in AGB Stars}\label{sec:AstroImpli_AGB}

\begin{figure}[htbp]
\epsscale{0.75}
\plotone{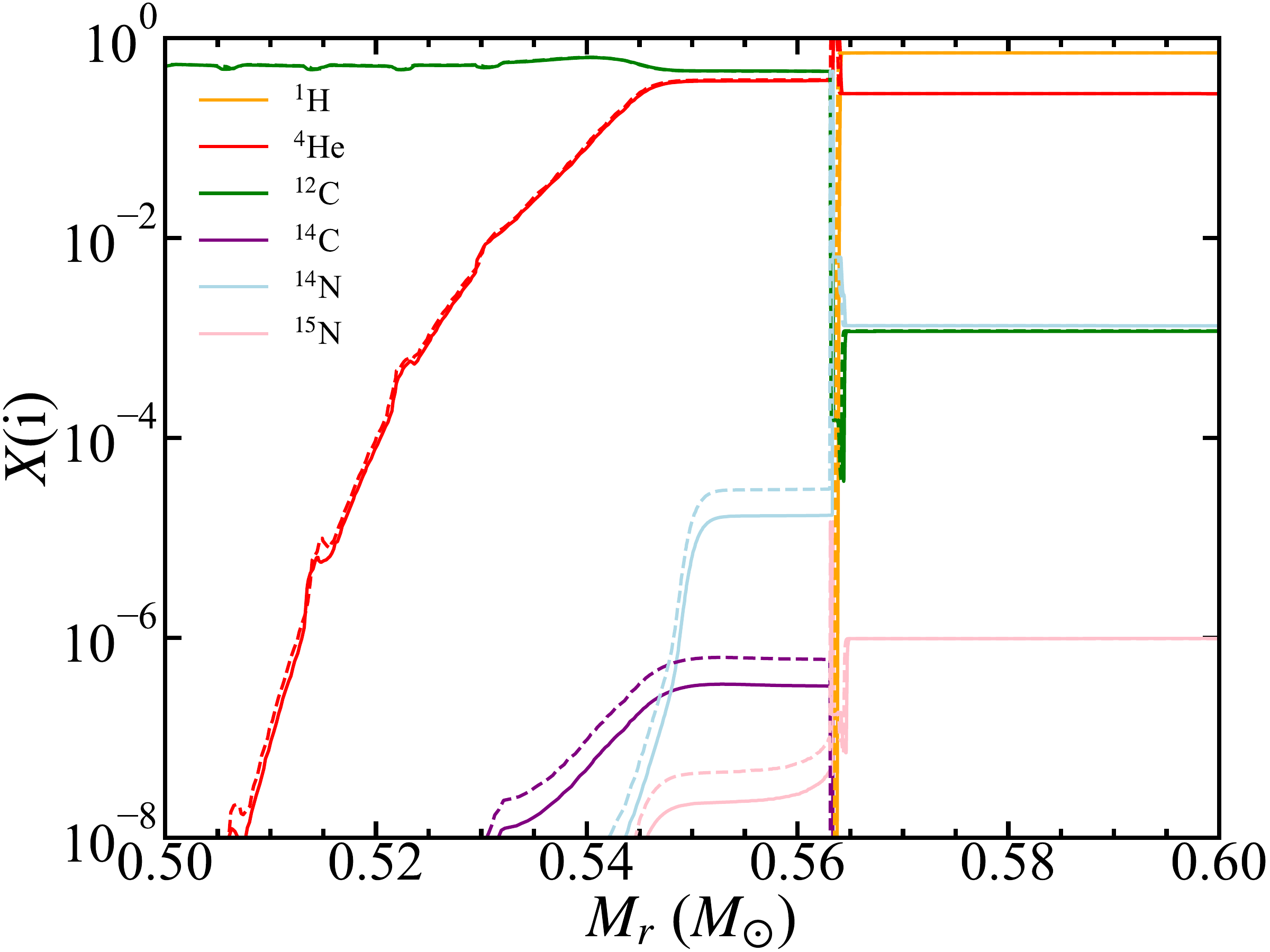}
\caption{The mass distribution of several isotopes from $M_r$ = 0.5 - 0.6 $M_\odot$, corresponding to the Helium shell and H-rich envelope. The model with different $^{14}$C (n, $\gamma$)$^{15}$C reaction rate is calculated after ten thermal pulses and a $^{13}$C pocket is formed near $M_r$ = 0.564 $M_\odot$. The solid and dashed lines represent the model with the reaction rate from JINA REACLIB and the rate in this work, respectively.}
\label{fig:agb_mr}
\end{figure}

In this section, we investigate the effect of our $^{14}$C(n, $\gamma$)$^{15}$C reaction rate on the nucleosynthesis in AGB stars. We utilized the Modules for Experiments in Stellar Astrophysics (MESA, version 12778;~\citealt{2011ApJS..192....3P, 2015ApJS..220...15P, 2018ApJS..234...34P, 2023ApJS..265...15J}) to follow the structural evolution and nucleosynthesis of hydrogen and helium burning with an initial mass of 2 $M_\odot$ and an initial metal abundance of 0.5 $Z_\odot$. We modified the example in the MESA directory, specifically ``agb\_to\_wd'', to create an AGB star model and track its evolution through more than ten thermal pulses. The $^{13}$C pocket forms in a thin layer near $M_r$ = 0.564 $M_\odot$, where $M_r$ is the Lagrangian mass-coordinate in units of the solar mass $M_\odot$ as shown in Figure~\ref{fig:agb_mr}. The abundance distributions of several isotopes using the $^{14}$C(n, $\gamma$)$^{15}$C reaction rate from the JINA REACLIB and our new rate are shown separately in Figure~\ref{fig:agb_mr}. Motivated by the significance of the mass fraction between $^{14}$N and $^{15}$N, i.e., $^{14}$N/$^{15}$N, which acts as one of the most important tracers of stellar nucleosynthesis and metal enrichment in the interstellar medium (see, e.g.,~\citet{Chen_ApJ_Interstellar_2024,Pignatari_ApJL_CARBONRICH_2015,Hedrosa_ApJL_NITROGEN_2013}), and the simple fact that $^{14}$C and $^{15}$C convert to $^{14}$N and $^{15}$N through $\beta^{-}$ decay respectively, we investigated the evolution of the abundances of $^{14}$N and $^{15}$N.

Using our new rate, the mass fraction of $^{14}$C in the layers 0.53 $M_\odot$ $\leq$ $M_{r}$ $\leq$ 0.564 $M_\odot$ increases by a factor of 1.84. This enhancement is attributed to the differences in the $^{14}$C(n, $\gamma$)$^{15}$C reaction rate between this work and the JINA REACLIB database (see Figure~\ref{fig-ReactionRate}). The elevated abundance of $^{14}$C subsequently influences the nitrogen isotopic composition in this region through two distinct processes: (1) the $\beta^{-}$ decay of $^{14}$C leads to an increase of the abundance of $^{14}$N by a factor of 1.83, and (2) the subsequent neutron capture on $^{14}$N results in an increase of the abundance of $^{15}$N by a factor of 2.00, as shown in Figure~\ref{fig:agb_mr}.

However, such a slight change in the inner region hardly affects the chemical composition in the interstellar medium of the Galaxy because there is almost no $^{14}$C in the hydrogen envelope ($M_r \geq$0.56 $M_\odot$) and the surface abundances of $^{14}$N and $^{15}$N originate primarily from the initial abundances~\citep{1995ApJS..101..181W}. If the materials are transported from the inner region to the surface through the dredge-up or some other efficient mixing and ejection processes, pre-solar grains that originate from the ejecta of AGB stars might indicate the variance of the abundance ratios among $^{14}$C (with half-life $T_{1/2} \approx$ 5700 yr), $^{14}$N and $^{15}$N if they form after a few years but within 30 yr as the Supernova (SN) grains could do~\citep{Liu_Sci.Adv._Late_2018}. 
The boundary of the carbon-oxygen (CO) core is defined at $M_r$ = 0.52 $M_\odot$, where the mass fraction of He decreases to $10^{-4}$~\citep{Nomoto2023, 2025arXiv250211012X}. When this AGB star ends its life, a CO white dwarf (WD) is formed, and the outer layers of the CO core are blown into space. However, compared to the total mass of the outer layer of 1.48 $M_\odot$, the mass changes occurring between 0.53 $M_\odot$ $\leq$ $M_r$ $\leq$ 0.56 $M_\odot$ will not have a big effect after integration.

\subsection{Inhomogeneous Big-Bang Nucleosynthesis}\label{sec:AstroImpli_IBBN}

After the cosmic QCD phase transition, baryons could form inhomogeneously with both low- and high-density regions coexisting in the early Universe. As the cosmic temperature drops to $\sim 1$ MeV, weak interactions decouple, and neutrons could diffuse out almost freely from high-density to low-density regions. On the other hand, protons have a large scattering cross section between them and the background electron-positron plasma~\citep{Applegate:1985qt,Applegate:1987hm}. Due to such different diffusion scales of protons and neutrons, one can expect that a relatively neutron-rich region could be formed, and the primordial nucleosynthesis there will be different from the standard BBN~\citep{Fuller:1987ue,Alcock:1987tx,1990ApJ...358...36M}. In the neutron-rich region, heavy elements could be produced through the enhanced reaction chain $^{1}$H(n, $\gamma$)$^{2}$H(n, $\gamma$)$^{3}$H(d, n)$^{4}$He(t, $\gamma$)$^7$Li(n, $\gamma$)$^8$Li($\alpha$, n)$^{11}$B(n, $\gamma$)$^{12}$B($\beta^{-}$)$^{12}$C(n, $\gamma$)$^{13}$C(n, $\gamma$)$^{14}$C. Then, as pointed out in the previous study~\citep{Kajino_PrimordialNucleosynthesisIntermediateMass_1990a}, the $^{14}$C(n, $\gamma$)$^{15}$C reaction could help form the heavier elements.

In this section, we investigate the difference that the newly measured $^{14}$C(n, $\gamma$)$^{15}$C reaction rate could make in the IBBN framework. Based on the coupled diffusion and nucleosynthesis code~\citep{1990ApJ...358...36M}, a condensed sphere model was applied for the baryon density inhomogeneity in this calculation. The baryon-to-photon ratio $\eta$ was set to be $6.13\times 10^{-10}$, taken from the final Planck result~\citep{Tristram:2023haj}. Three parameters are required to describe the baryon inhomogeneity: the density contrast $R$ between the high-density and low-density regions, the volume fraction $f_v$ of the high-density region, and the average separation distance $r$ between fluctuations. We used the optimized value $R = 10^6$ and $f_v^{1/3}=0.5$ from~\citet{Orito:1996zb}. 

\begin{figure}[htbp]
\epsscale{0.75}
\plotone{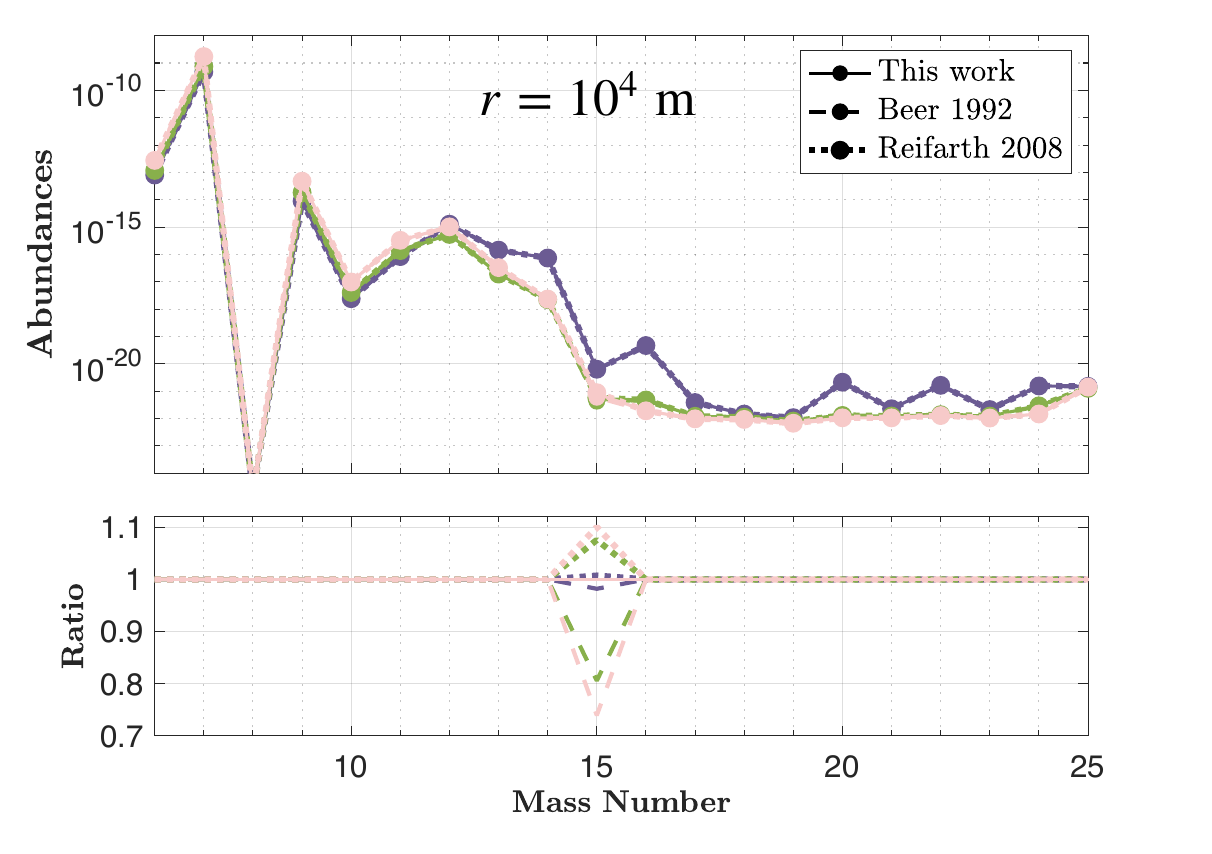}
\plotone{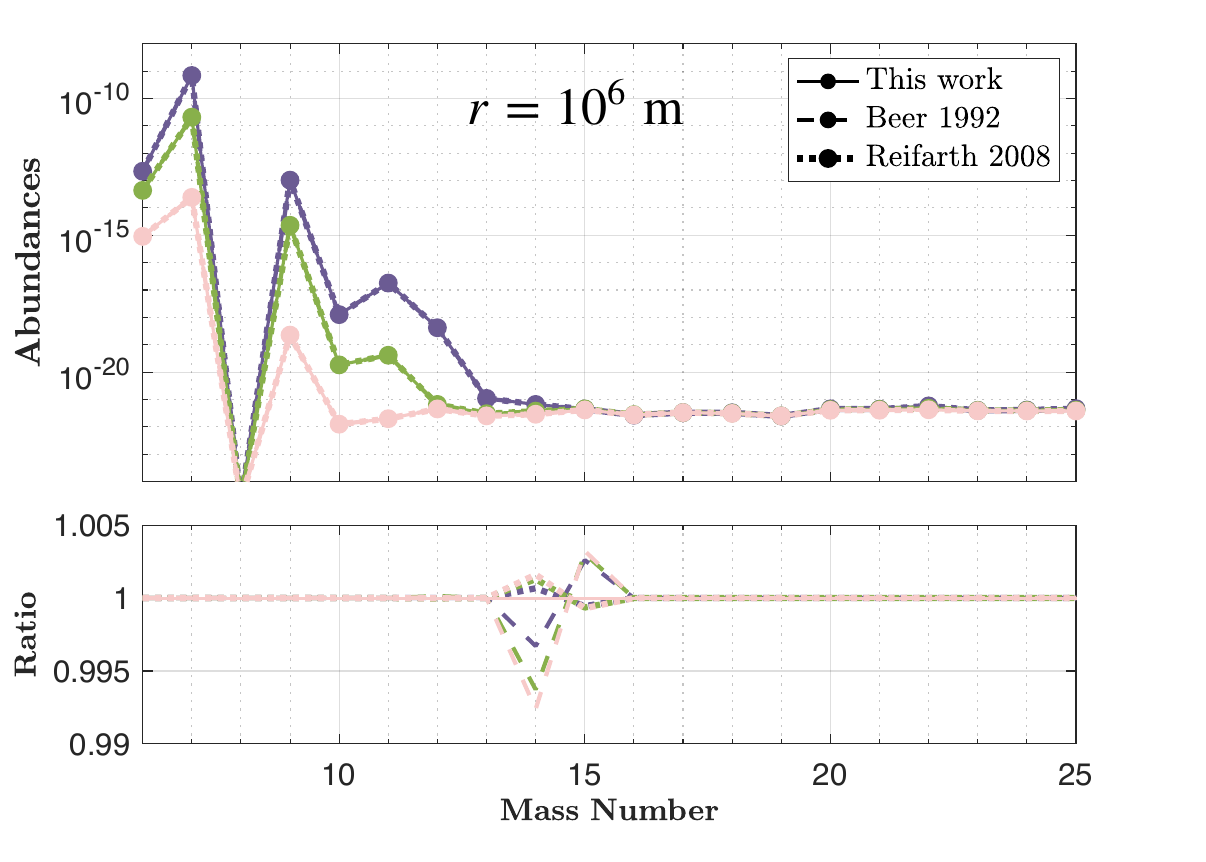}
\caption{Top panel: The final abundances of an IBBN model with $r=10^4$ m. We show the case of  Zone No.13(pink lines), No.14(green lines), and No.15(purple lines) in our calculation, where the neutron-rich condition is valid so that nuclei with A$>14$ are produced. The ratios of the final abundances using our new rate to the previous rates are also shown below in the abundance plot. Bottom panel: Same as the top panel but with separation distance $r=10^6$ m. In this scenario, Zone 13$\sim$15 are far from high density region, neutron injection effect is tiny, and the reaction flow did not pass $^{14}$C(n, $\gamma$)$^{15}$C, so the new measured reaction rate did not make a significant difference.}
\label{fig:IBBN}
\end{figure}

The neutron-rich condition is valid near the boundary between high-density and low-density regimes.
To explore $^{14}$C(n, $\gamma$)$^{15}$C effect in these regions, we set $r=10^4$ m in IBBN code and choose zone No. 13$\sim$15 (there are 16 zones in total) as an example\footnote{In this study, we scale $r$ value at the epoch of cosmic temperature $T = 1$ MeV.}. The results of the final abundances using our new rate, the Beer 1992 rate, and the Reifarth 2008 rate are represented by solid, dashed, and dotted lines in the top panel of Figure~\ref{fig:IBBN}, respectively. For the typical temperature of BBN ($T_{9}$ $\approx$ 0.5), the Beer 1992 rate is about 3.4 times lower than our new rate, while Reifarth 2008 is about 30\% higher (see Figure~\ref{fig-ReactionRate}). In these neutron-rich zones, $^{14}$C(n, $\gamma$)$^{15}$C reaction is the main reaction to produce $A=15$ elements. Therefore, the final abundance of $^{15}$N is approximately 20\% lower if using the Beer 1992 rate and 10\% higher if using the Reifarth 2008 rate in these regions.

When the separation distance $r$ becomes larger, neutron injection could not reach the low-density zones that far from the high-density regions, so that low-density zones are almost the same as that in the standard BBN (i.e., still proton-rich), but with a much lower baryon density. We also investigate the role of $^{14}$C(n, $\gamma$)$^{15}$C in this situation: by setting $r=10^6$ m, the above zone No. 13$\sim$15 become such cases. This time, the nuclei with $A>14$ are not significantly synthesized, and the reaction flow did not pass $^{14}$C(n, $\gamma$)$^{15}$C, thus the primary production reaction of the $A=15$ elements is $^{14}$C(p, $\gamma$)$^{15}$N. Hence, the reduction of the (n, $\gamma$) reaction rate makes the flow toward (p, $\gamma$) slightly easier and vice versa, leading to about 0.5\% change in the $A=14$ and $A=15$ nuclear systems in the low-density regions.

\subsection{\textit{r}-process in core-collapse supernovae}~\label{sec:AstroImpli_Rprocess}

The \textit{r}-process is responsible for the synthesis of nearly half of the elements heavier than iron in the universe. Since the detection of the gravitational wave GW170817 and its associated kilonova GRB170817~\citep{abbott2017gravitational,abbott2017gw170817}, the neutron star merger (NSM) as well as core-collapse supernova (CCSN) have been confirmed to be significant astrophysical sites for the \textit{r}-process. However, NSM has an unavoidable long time-delay of the cosmological timescale, which makes it difficult to explain the enrichment of \textit{r}-process elements in the early universe~\citep{Cote_Astrophys.J._Neutron_2019,Yamazaki_ApJ_Possibility_2022,Kobayashi_ApJL_Can_2023}. The magneto-hydrodynamically driven jet (MHDJ) from rapidly rotating magnetized CCSNe~\citep{winteler2012,nishimura2015} and the collapsars, which leave black holes instead of neutron stars as remnants of the explosion of very massive progenitor stars~\citep{Yamazaki_ApJ_Possibility_2022, He_ApJL_Possibility_2024}, are the promising sites to be the dominant contributors to the \textit{r}-process abundances in the metal-pool epoch of the early universe~\citep{Kajino_ProgressinParticleandNuclearPhysics_Current_2019}. 

The \textit{r}-process in MHDJ and collapsars starts under the conditions of high temperatures and high neutron densities. Initially, the temperature is high enough around $T$ $\sim$ 10~GK, when the system is in nuclear statistical equilibrium (NSE), consisting primarily of protons and neutrons with a small amount of $\alpha$-particles. As the temperature drops to $T$ $\approx$ 5~GK, NSE is no longer satisfied and a large number of seed elements like $^{78}$Ni are produced by the $\alpha$-process. When the temperature decreases further to $T$ $\approx$ 5/e $\approx$ 2~GK, charged-particle induced reactions almost freeze out ($\alpha$-rich freezeout) due to the Coulomb barrier. Then, neutron captures followed by $\beta^{-}$ decays dominate the \textit{r}-process.

\begin{figure}[htbp]
\epsscale{0.75}
\plotone{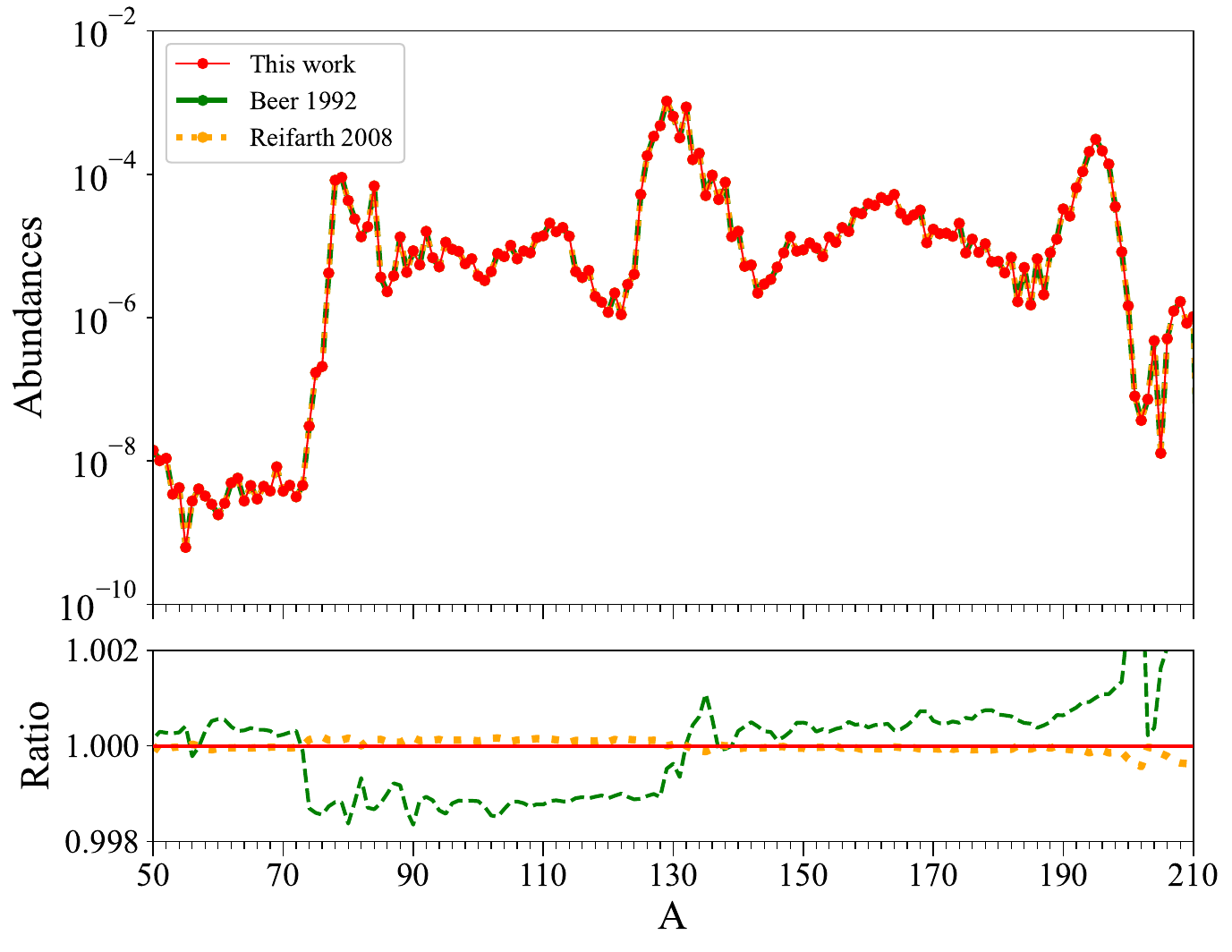}
\caption{Top panel: Mass-weighted sum of final abundances of the \textit{r}-process nucleosynthesis in the MHDJ. The red, green, and orange curves correspond to the cases using our new rate, the Beer 1992 rate, and Reifarth 2008 rate, respectively; Bottom panel: Ratios of the final abundances obtained with the previous reaction rate to those obtained using our updated rate.}
\label{fig:ya_mhd}
\end{figure}

The early studies of the \textit{r}-process in neutrino-driven winds (NDWs) from CCSNe~\citep{terasawa2001new, Sasaqui_SensitivityProcessNucleosynthesis_2005b} indicate that nuclear reaction paths in the light-mass neutron-rich region can change the final abundances of the \textit{r}-process elements by an order of magnitude. Motivated by these studies, we extensively investigated the impact of the new rate of $^{14}$C(n, $\gamma$)$^{15}$C on the \textit{r}-process nucleosynthesis in MHDJ based on the codes of NucNet Tools~\citep{meyer2007}. We adopted 23 trajectories to calculate \textit{r}-process nucleosynthesis from the hydrodynamical simulation of MHDJ~\citep{nishimura2015}. The impact of our new $^{14}$C(n, $\gamma$)$^{15}$C rate on the final abundances is illustrated in Figure~\ref{fig:ya_mhd}, where, similar to the investigations in the previous subsections, comparisons of our result with nucleosynthesis using the $^{14}$C(n,$\gamma)^{15}$C rates from~\citet{Beer_Measurement14CGamma_1992} and~\citet{Reifarth_14CrossSection_2008} were also made.

\begin{figure}[htbp]
\centering
\includegraphics[width=0.9\textwidth]{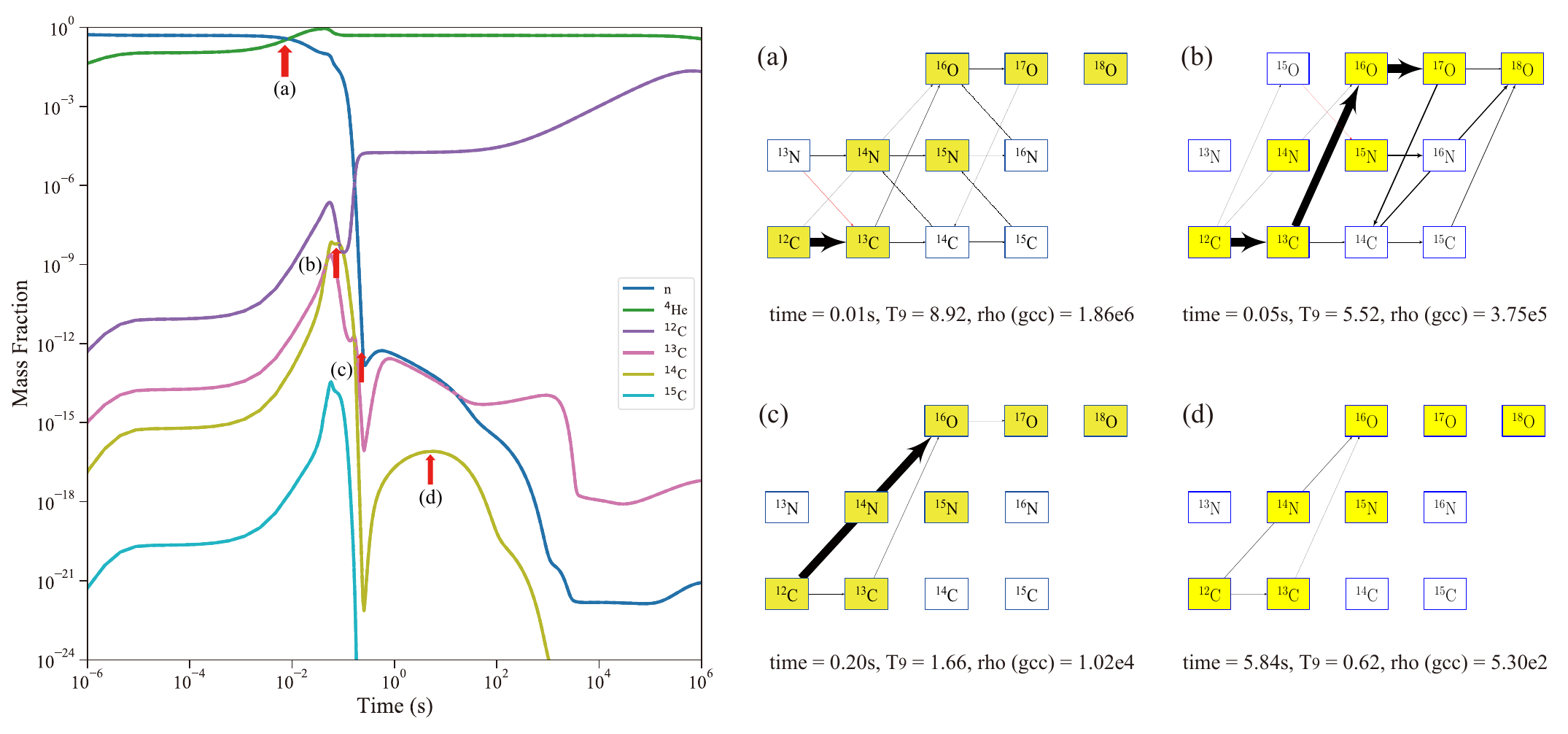}
\caption{Left panel: Evolution of the mass fractions for neutrons, $^{4}$He, and typical carbon isotopes. Right panel (a-d): Reaction flows in the CNO region at $\sim 0.01$ s, $\sim 0.05$ s, $\sim 0.20$ s, and $\sim 5.84$ s, as indicated by the red arrows in the left panel. Note that the width of each arrow is proportional to the magnitude of the reaction flow $(dY_{i}/dt)_{j}$, which is defined as the rate of abundance change of nuclide $i$ via reaction $j$.}
\label{fig:flow_path}
\end{figure}

We found that the differences between the final abundances using the Beer 1992 rate and our new rate fluctuate by only around $\sim$ 0.2\%, while the differences between results using the Reifarth 2008 rate and our rate are nearly negligible. This is reasonable because the difference between the Beer 1992 rate and ours is much larger than that between the Reifarth 2008 rate and ours. We note that the Beer 1992 rate is lower than our updated rate throughout the temperature range, which implies that fewer neutrons are consumed during the $\alpha$-process to produce seed nuclei (see Figure~\ref{fig-ReactionRate}). As a result, it leads to slightly higher abundances of heavy elements with $A > 130$ when the Beer 1992 rate is adopted. To understand the role of $^{14}$C(n, $\gamma$)$^{15}$C in our nucleosynthesis calculations, we studied the time evolution of neutron, $^{4}$He, and carbon isotopes $^{12-15}$C, as shown in Figure~\ref{fig:flow_path}.

In the evolution of the carbon isotopic mass fractions in the left panel of Figure~\ref{fig:flow_path}, we found that $^{14}$C (yellow line) is abundant before the neutron freezeout at $t$~$\sim 0.20$~s, but rapidly diminishes afterward. Although the mass fraction of $^{14}$C increases again after $\sim 1$ s, it remains negligible compared to those of $^{12}$C and $^{13}$C. We selected four typical moments during the nucleosynthesis, as marked by the red arrows: (a) $0.01$~s when $^{4}$He is predominantly produced through the fusion of free neutrons and protons; (b) $0.05$~s when $^{14}$C reaches its peak mass fraction before the neutron freezeout; (c) $0.20$~s when neutron is nearly exhausted; (d) $5.84$~s, when $^{14}$C reaches its peak mass fraction after the neutron freezeout. Then, we investigated the reaction flows in the CNO mass region at the four stages adopted above. Before the mass fraction of $^{4}$He surpasses that of neutrons, as shown in Figure~\ref{fig:flow_path} (a), the neutron-induced reaction $^{12}$C(n, $\gamma$)$^{13}$C dominates the reaction flows in the CNO mass region. However, once $^{4}$He becomes the most abundant nucleus, the $\alpha$-induced reaction $^{13}$C($\alpha$, n)$^{16}$O becomes as significant as the neutron capture reactions on $^{12}$C and $^{16}$O (Figure~\ref{fig:flow_path} (b)). After the neutron freezeout shown in Figure~\ref{fig:flow_path} (c), the $^{12}$C($\alpha$, $\gamma$)$^{16}$O reaction dominates, while (n, $\gamma$) reactions become negligible. Although the mass fraction of $^{14}$C increases again later, the dominant reaction paths during this period reside in the region consisting of heavy nuclei, making the reactions in the CNO mass region relatively insignificant (Figure~\ref{fig:flow_path} (d)). In this sense, the $^{14}$C(n, $\gamma$)$^{15}$C reaction does not lie on the main reaction path during nucleosynthesis, which explains why only minor differences of $\sim 0.2\%$ can be seen in the final abundance pattern displayed in Figure~\ref{fig:ya_mhd}. This may contradict the early theoretical studies proposed in~\citet{terasawa2001new,Sasaqui_SensitivityProcessNucleosynthesis_2005b} who used the NDW models of CCSNe assuming very low initial electron-mole fraction $Y_e$, or equivalently very high initial n/p-ratio, and relatively high entropy-per-baryon~\citep{Meyer_Astrophys.J._RProcess_1992, Woosley_Astrophys.J._RProcess_1994}, but it is understandable because later theoretical studies of the neutrino transport during core collapse and explosion~\citep{Fischer_A&A_Protoneutron_2010} and of the mean-field effects on neutrino interactions at high densities~\citep{Roberts_Phys.Rev.C_Medium_2012} revealed that these appropriate conditions for successful \textit{r}-process are not necessarily available in NDWs from CCSNe. We have confirmed that the $^{14}$C(n, $\gamma$)$^{15}$C reaction rate is now determined precisely enough such that the uncertainty does not affect the \textit{r}-process yields in the MHDJ model which is one of the viable astrophysical sites for the \textit{r}-process~\citep{Kajino_ProgressinParticleandNuclearPhysics_Current_2019, Yamazaki_ApJ_Possibility_2022}. This is consistent with a recent systematic sensitivity study of the light-mass nuclear reactions up to oxygen isotopes in the \textit{r}-process nucleosynthesis~\citep{Kim_Sensitivity_2025} where successful MHDJ and collapsar models as the one adopted in this article is used.


\section{Summary and Conclusion} \label{sec:conclusion}

In this paper, we obtain the SF and ANC for $^{15}$C$_{\mathrm{g.s.}}$ for the first time from the perspective of neutron $removal$ transfer reactions. The SF and ANC have been used to determine the $^{14}$C(n, $\gamma$)$^{15}$C cross section at astrophysical low energies and the reaction rate. Within the temperature range of 0.01-4 GK, our new reaction rate is 2.4-3.7 times higher than that deduced from the first direct measurement~\citep{Beer_Measurement14CGamma_1992} and 20\%-25\% lower than that obtained in the most recent direct measurement~\citep{Reifarth_14CrossSection_2008}. These results effectively complement the results deduced from previously studied nuclear $addition$ transfer reactions, thus paving the way for systematic research on this kind of (n, $\gamma$) reactions. Thanks to the availability of complementary data obtained from studies such as Coulomb dissociation and charge symmetry, more extensive efforts such as global analysis based on the multi-channel \textit{R}-matrix method (see, e.g.,~\citet{Lane_Rev.Mod.Phys._RMatrix_1958,DeBoer_Rev.Mod.Phys._C12O16Reaction_2017,Nan_PhysicsLettersB_Determination_2025}) will illustrate a more comprehensive picture of the nature of such (n, $\gamma$) reactions in the future.

Considering the numerous challenges and uncertainties associated with direct measurements of (n, $\gamma$) reactions of astrophysical interest, this work demonstrates the feasibility and significance of conducting nuclear astrophysics research at modern rare isotope facilities, especially for nuclear reactions that involve nuclei far from the valley of stability, where direct measurements are currently unfeasible with existing technology. Furthermore, this work indicates that 
by deepening our understanding of nuclear physics, such as the ``quenching'' effect, our comprehension of significant nuclear reactions in cosmic and stellar evolution will be greatly improved. Consequently, it provides a broader perspective for us to address several cosmological and astrophysical problems in future studies.

Motivated by early theoretical studies on this reaction, we have further explored nucleosynthesis in AGB stars, inhomogeneous Big-Bang nucleosynthesis, and the \textit{r}-process using state-of-the-art theoretical models available today. {We found that the abundances of $^{14}$N and $^{15}$N can be enhanced by factors of 1.83 and 2.00, respectively, in the inner regions of AGB stars. Although this enhancement has a negligible impact on the overall chemical composition in the interstellar medium due to the relatively small mass of the inner region of 0.02 $M_{\odot}$ compared to the total mass of the outer layers of 1.48 $M_{\odot}$, the variation in the $^{14}$N/$^{15}$N ratio can be traced through pre-solar grains originating from AGB star ejecta.
In the context of inhomogeneous Big Bang nucleosynthesis, our new reaction rate for $^{14}$C(n, $\gamma$)$^{15}$C can lead to a difference up to $\sim 20\%$ in the final abundance of $^{15}$N within the boundaries between high-density and low-density regions where neutron injection occurs. As for the \textit{r}-process, our calculations indicate that our newly determined $^{14}$C(n, $\gamma$)$^{15}$C reaction rate only results in a difference of approximately $\sim 0.2\%$ in the final abundances of heavy elements with $A > 90$ in the MHDJ model. This confirms that our newly measured reaction rate is sufficiently precise to ensure that its uncertainty does not significantly impact the $r$-process yields in the MHDJ model.}


Our results will stimulate more new work, particularly systematic studies of nuclear astrophysical reactions involving unstable nuclei. Thanks to the numerous ongoing experiments at platforms such as the ATLAS facility mentioned in this paper, the Beijing Radioactive Ion beam Facility (BRIF), the Isotope Separator On Line DEvice (ISOLDE), and the Facility for Rare Isotope Beams (FRIB), we are optimistic that such systematic studies will be carried out in the near future.





\section{Acknowledgments}
This work was supported by the National Key R\&D Program of China (Grant No. 2022YFA1602301, 2022YFA1602401), and the National Natural Science Foundation of China (Grants No. 12435010, No. 11490560, No. 12125509, No. 12222514, No. U2441220, No. 12335009). The authors thank Nan Liu for constructive discussions on meteorite science. The authors thank Xiaodong Tang for suggestions and discussions on radiative capture reaction theory.

\bibliography{14CngDraft}{}
\bibliographystyle{aasjournal}



\end{document}